\documentclass[reprint,prapplied,amsmath,amssymb,aps,showkeys,floatfix,superscriptaddress]{revtex4-2}
\usepackage[utf8]{inputenc}
\usepackage[T1]{fontenc}
\usepackage[english]{babel}
\usepackage{graphicx}
\usepackage{siunitx}
\usepackage{amsmath,amssymb}
\usepackage{ulem}
\usepackage{lipsum}
\usepackage{hyperref}
\usepackage{soul}
\usepackage{comment}

\begin{document}

\title{Pulse-resolved post-generation phase control of squeezed light at 93 MHz with a bulk lithium-niobate modulator}

\author{Edoardo Suerra}
\email{edoardo.suerra@unimi.it}
\affiliation{Dipartimento di Fisica, Università degli Studi di Milano, Via Celoria 16, Milan, Italy}

\author{Igor Vasiljevic}
\affiliation{Dipartimento di Fisica, Università degli Studi di Milano, Via Celoria 16, Milan, Italy}
\affiliation{Istituto Nazionale di Fisica Nucleare, Sezione di Milano, Via Celoria 16, Milan, Italy}

\author{Sebastiano Corli}
\affiliation{Dipartimento di Fisica, Università degli Studi di Milano, Via Celoria 16, Milan, Italy}
\affiliation{Istituto di Fotonica e Nanotecnologie - CNR, Piazza Leonardo da Vinci 32, Milan, Italy}

\author{Enrico Prati}
\affiliation{Dipartimento di Fisica, Università degli Studi di Milano, Via Celoria 16, Milan, Italy}
\affiliation{Istituto Nazionale di Fisica Nucleare, Sezione di Milano, Via Celoria 16, Milan, Italy}
\affiliation{Istituto di Fotonica e Nanotecnologie - CNR, Piazza Leonardo da Vinci 32, Milan, Italy}

\author{Simone Cialdi}
\affiliation{Dipartimento di Fisica, Università degli Studi di Milano, Via Celoria 16, Milan, Italy}
\affiliation{Istituto Nazionale di Fisica Nucleare, Sezione di Milano, Via Celoria 16, Milan, Italy}

\begin{abstract}
Time-domain multiplexing of continuous-variable quantum states provides a scalable approach to photonic quantum information processing, in which individual temporal modes can be sequentially manipulated within a common optical path. 
Here we demonstrate pulse-resolved post-generation phase control of squeezed-vacuum states at a repetition rate of \SI{93}{\mega\hertz} using a low-loss bulk lithium-niobate electro-optic modulator driven by a broadband RF power amplifier. 
The driving architecture exploits low-cost, commercially available RF components and the impedance mismatch between the RF chain and the capacitive crystal load, rather than relying on dedicated custom high-voltage electronics, to obtain large voltage excursions on the modulator while operating above the piezoelectric-resonance region of lithium niobate.
The modulator is driven synchronously with the optical pulse train, enabling the phase of consecutive temporal modes to be individually controlled. 
An interferometric calibration yields a maximum differential phase excursion of \SI{1.93}{\radian}. 
We apply the modulator to a \SI{1035}{\nano\meter} squeezed-vacuum field generated by a synchronously pumped optical parametric oscillator and characterize the output using time-resolved balanced homodyne detection. 
With the local-oscillator phase kept fixed, the phase of the squeezed state is varied from pulse to pulse, and the corresponding phase-dependent quadrature variance is reconstructed. 
The resulting quadrature scan agrees with an equivalent measurement obtained by varying the local-oscillator phase with a piezoelectric actuator. 
These results demonstrate post-generation quantum-state phase control at the individual-pulse level and provide an electro-optic building block for programmable manipulation of time-multiplexed quantum states in recirculating optical architectures.
\end{abstract}

\maketitle

\section{Introduction}\label{sec:introduction}
Optical systems are promising platforms for quantum information processing (QIP), owing to their long coherence times, room-temperature operation, and potential for high repetition rates, with applications spanning quantum computing, quantum sensing, and quantum communication~\cite{Flamini2019,Pirandola2018,Lawrie2019}.
In particular, combining continuous-variable quantum states with time-domain multiplexing provides a scalable approach to optical QIP, in which quantum information is encoded into a sequence of temporally separated optical modes propagating through a limited number of spatial modes~\cite{Takeda2019}.
Time-multiplexed squeezed-light sources have enabled the generation of large-scale entanglement and the implementation of quantum information protocols based on both measurement-based and loop-based architectures~\cite{Yokoyama2013,Motes2014,Yoshikawa2016,Larsen2019,Asavanant2019,Larsen2021,Asavanant2021,Takeda2017,Takeda2019_2,Enomoto2021}.

A key requirement of time-multiplexed architectures is the ability to manipulate individual temporal modes of the optical field.
In general, different pulses may need to undergo different quantum operations depending on their position within the temporal sequence.
This requirement is particularly relevant in loop-based architectures, where successive temporal modes repeatedly traverse the same optical components and programmable pulse-resolved operations determine the evolution of the quantum state during propagation through the loop~\cite{Takeda2017,Takeda2019_2,Enomoto2021,Yonezu2023}.
The possibility of controlling the phase of individual temporal modes is especially important, since a phase shift corresponds to a rotation of the optical quadratures and therefore constitutes a fundamental operation on continuous-variable quantum states.

Recent experiments have demonstrated programmable control of time-multiplexed squeezed light by modulating the pump field of an optical parametric amplifier, allowing the squeezing level and phase to be programmed on a pulse-by-pulse basis~\cite{Tomoda2023}.
In this approach, the desired temporal pattern is established \textit{during} the generation of the squeezed states.
An alternative and complementary approach is to apply the optical transformation \textit{after} the quantum state has already been generated.
Post-generation control is particularly attractive for architectures in which a common quantum source feeds a sequence of dynamically programmable operations, as it allows the same generated state to be selectively transformed at different positions of the optical circuit.
When implemented inside a recirculating optical path, such an operation can provide a compact mechanism for dynamically controlling successive temporal modes while separating quantum-state generation from subsequent temporal-mode-dependent manipulation and avoiding additional switching stages.

Electro-optic modulation provides a direct route to high-speed optical phase control through the linear electro-optic effect. 
Lithium niobate (LiNbO$_3$, LN) is particularly attractive for this purpose because of its strong electro-optic response, broad optical transparency, and established use in high-speed optical modulators~\cite{Weis1985,Zhang2021}.
While integrated lithium-niobate modulators can provide electro-optic bandwidths exceeding $\SI{100}{\giga\hertz}$ in compact footprints~\cite{Zhang2021}, bulk devices offer an alternative particularly suited to low-loss free-space optical systems, where preserving the transmission of nonclassical states is essential.
Bulk electro-optic modulators have therefore also been employed for programmable manipulation of time-domain quantum states~\cite{Carosini2024,Madsen2022}.
However, high-speed operation of bulk lithium-niobate modulators poses an additional challenge due to the electromechanical response of the crystal. 
Variations of the applied electric field can excite piezoelectric modes, producing a mechanical contribution to the optical phase through both changes in the optical path length and the photoelastic effect~\cite{Takeda2012}. 
Broadband measurements of lithium niobate identify distinct frequency regimes associated with free-piezoelectric, clamped-ionic, and electronic responses, separated by resonances~\cite{Cochard2017}.
In particular, a pronounced piezoelectric-resonance region occurs in the vicinity of $\SI{10}{\mega\hertz}$, whereas at higher frequencies the response approaches the clamped-ionic regime. 
Operation in the latter regime therefore provides a way to suppress the mechanically mediated contribution while
retaining the direct electro-optic response.

Previous demonstrations of programmable phase control using bulk electro-optic devices have operated at substantially lower update rates. 
In Ref.~\cite{Carosini2024}, a free-space lithium-niobate phase modulator was reconfigured to an arbitrary phase value every \SI{100}{\nano\second}, corresponding to a maximum operating rate of \SI{10}{\mega\hertz}. 
The Borealis processor operated with a \SI{6}{\mega\hertz} pulse train and employed bulk electro-optic phase shifters for dynamic programming of the time-domain interferometer~\cite{Madsen2022}. 
Pulse-resolved phase control was also demonstrated in a phase-stabilized dual-loop processor with an update interval of \SI{66}{\nano\second}, corresponding to approximately \SI{15}{\mega\hertz}~\cite{Yonezu2023}. 
These implementations demonstrate the usefulness of bulk electro-optic modulation for time-domain quantum processing, but operate well below the \SI{93}{\mega\hertz} clock rate considered here and rely on dedicated modulator driving systems.

Here we demonstrate a different approach to high-speed bulk lithium-niobate phase modulation. 
Rather than employing a dedicated high-voltage modulator driver, we drive a commercially available bulk LN crystal using a broadband RF power amplifier and a programmable arbitrary waveform generator. 
The electrical driving chain is operated above the piezoelectric-resonance region of LN, while the impedance mismatch between the \SI{50}{\ohm} RF chain and the capacitive crystal load is deliberately retained to enhance the voltage excursion at the modulator. 
This approach allows large voltages with fast electrical transitions to be generated from a low-amplitude AWG waveform using commercially available RF components. 
We use this architecture to demonstrate pulse-resolved phase control at a \SI{93}{\mega\hertz} repetition rate, corresponding to a temporal separation of approximately \SI{10.75}{\nano\second} between consecutive optical pulses.

The modulator is characterized electrically and through an interferometric measurement of the voltage-to-phase response, yielding a maximum differential phase excursion of \SI{1.93}{\radian}. 
We then apply the device to a \SI{1035}{\nano\meter} squeezed-vacuum field generated by a synchronously pumped optical parametric oscillator. 
By keeping the local-oscillator phase fixed and varying the applied electrical waveform from pulse to pulse, we reconstruct the phase-dependent quadrature variance of the generated state. 
The result is compared with an equivalent quadrature scan obtained by changing the local-oscillator phase with a piezoelectric actuator. 
The agreement between the two measurements demonstrates direct post-generation phase control of the quantum state at the individual-pulse level.

The demonstrated combination of high repetition rate, low optical loss, pulse-resolved operation, and a simple driving architecture based on commercially available RF components provides an electro-optic building block for the manipulation of time-multiplexed quantum states. 
In particular, the bulk implementation and compatibility with free-space optical paths make the device suitable for recirculating optical architectures, where successive temporal modes can undergo programmable phase transformations while propagating through the same optical circuit. 
Although the present experiment does not implement the modulator inside a recirculating loop, it demonstrates the pulse-resolved quantum-state manipulation required for such an application.

The paper is organized as follows.
Section~\ref{sec:setup} describes the experimental setup and the electro-optic phase modulator.
Section~\ref{sec:characterization} presents the electrical and interferometric characterization of the device.
Section~\ref{sec:methods} describes the pulse sequence and data-analysis procedure used for the quantum-state measurements.
The experimental demonstration of pulse-resolved phase control is presented in Section~\ref{sec:results}, and Section~\ref{sec:conclusions} concludes the paper.

\section{Experimental setup}\label{sec:setup}
The experimental setup is designed to demonstrate phase control of a quantum optical field after its generation, while preserving the temporal structure of the $\SI{93}{\mega\hertz}$ pulse train.
The experiment therefore combines three main elements: a synchronously pumped squeezed-light source, a low-loss bulk lithium-niobate electro-optic phase modulator driven by a programmable electrical waveform, and a time-resolved balanced homodyne detector.
The phase modulator is placed between the squeezed-light source and the homodyne detector, so that the generated quantum state is modified only after leaving the nonlinear generation stage.
A synchronized trigger derived from the mode-locked laser provides the timing reference for both the electrical modulation and the temporal-mode-resolved homodyne measurement.
A schematic overview of the experimental arrangement is shown in Figure~\ref{fig:setup}.

\subsection{Bulk lithium-niobate phase modulator}\label{subsec:modulator}
\begin{figure}
    \centering
    \includegraphics[width=85mm]{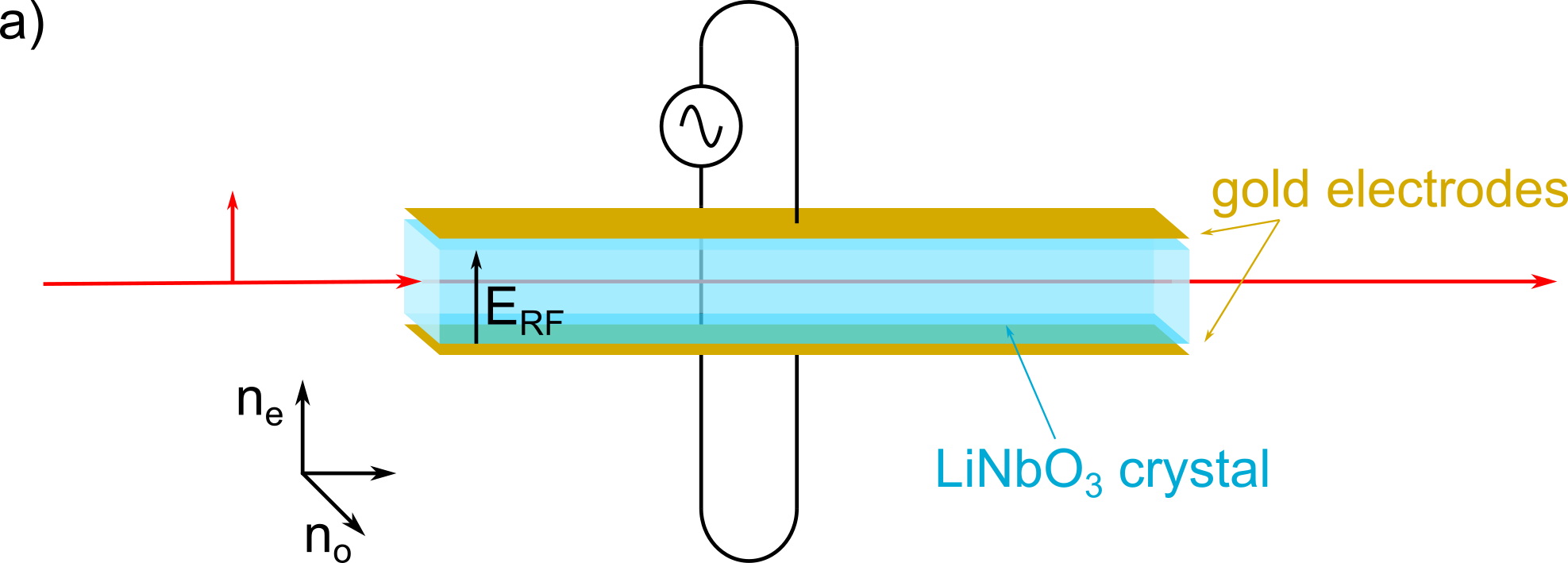}\\
    \vspace{4mm} 
    \includegraphics[width=85mm]{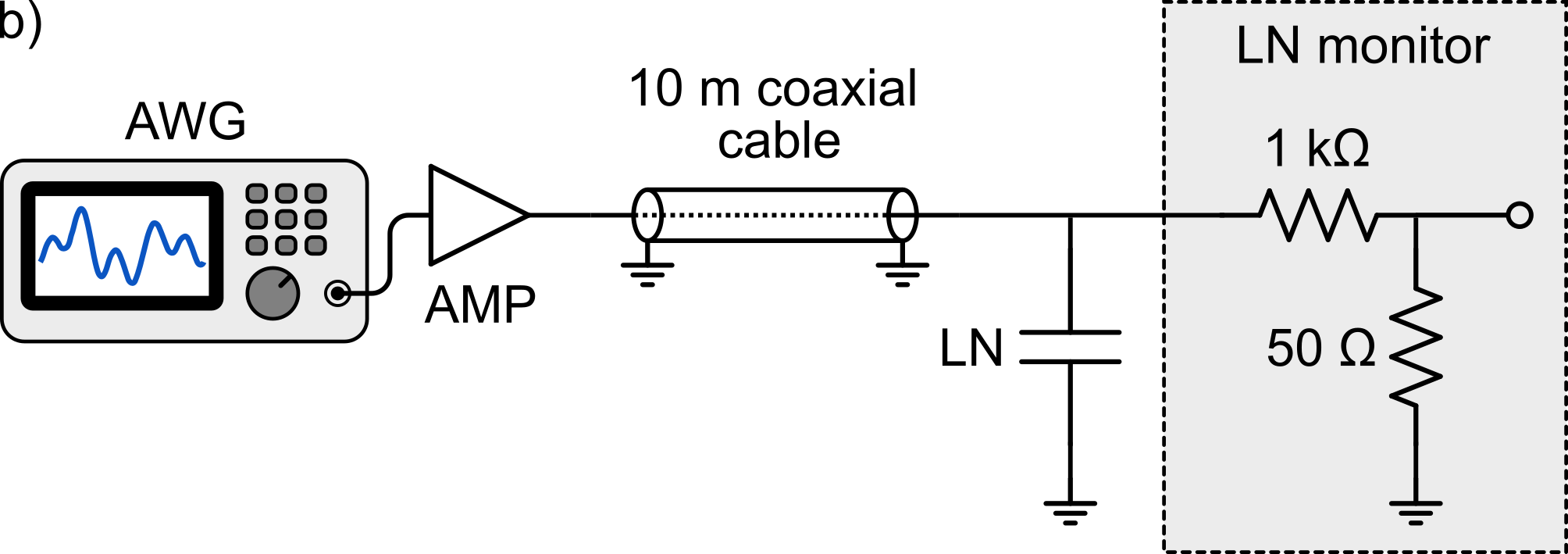} 
    \caption{
        Electro-optic phase modulator and electrical driving circuit. 
        (a) Geometry of the bulk lithium-niobate crystal. 
        Gold electrodes deposited on opposite surfaces generate an RF electric field, $\mathbf{E}_{\mathrm{RF}}$, oriented along the extraordinary crystallographic axis. 
        The optical field propagates orthogonally to $\mathbf{E}_{\mathrm{RF}}$ and is polarized along the extraordinary direction. 
        (b) Electrical circuit used to drive and monitor the crystal. 
        The waveform generated by the arbitrary waveform generator (AWG) is amplified and transmitted to the crystal through a $\SI{10}{\meter}$ coaxial cable. 
        A resistive voltage divider connected at the crystal input provides a monitor of the applied electrical waveform. 
    }
    \label{fig:LN}
\end{figure}

The phase modulator is based on a bulk MgO-doped lithium-niobate (LiNbO$_3$) crystal with a length of $\SI{40}{\milli\meter}$ and a cross section of $\SI{3}{\milli\meter}\times\SI{3}{\milli\meter}$, and it is schematized in Figure~\ref{fig:LN}a.
The crystal is anti-reflection coated for operation over the \SIrange{900}{1250}{\nano\meter} spectral range and introduces an optical loss of approximately \SI{3}{\percent} at the operating wavelength of \SI{1035}{\nano\meter}.
Two gold electrodes deposited on opposite crystal surfaces are used to apply the electric field for electro-optic modulation.

The optical field propagates through the crystal with its polarization oriented along the extraordinary optical axis.
The applied electric field therefore modifies the extraordinary refractive index through the linear electro-optic (Pockels) effect~\cite{Li2024}, resulting in a voltage-dependent optical phase shift. 
For an interaction length $L$, the induced phase variation can
be written as~\cite{Saleh1991_ch18}
\begin{equation}
    \Delta\phi(V)
    =
    \frac{2\pi L}{\lambda}\Delta n_e(V),
    \label{eq:phase_modulation}
\end{equation}
where $\lambda$ is the optical wavelength and $\Delta n_e(V)$ is the voltage-dependent variation of the extraordinary refractive index.
In the linear electro-optic regime, $\Delta n_e$ is proportional to the applied electric field and, for a fixed electrode geometry, to the applied voltage. 
The resulting voltage-to-phase conversion is therefore linear and is calibrated experimentally, as described in Section~\ref{subsec:phase_calibration}.

Following the electrical scheme of Figure~\ref{fig:LN}b, the modulator is driven using an amplified arbitrary waveform generator (AWG) from Active Technologies (model Arb Rider AWG-4000), with a sampling rate of \SI{1.2}{\giga S\per\second}.
The AWG is synchronized with the mode-locked laser, ensuring a fixed temporal relationship between the electrical waveform and the optical pulse train. 
The AWG output is then amplified by a broadband \SI{50}{\watt} RF power amplifier (AMP) from Mini-Circuits (model LZY-1), with a nominal gain of \SI{42}{\decibel} over the frequency range \SI{10}{\mega\hertz}--\SI{525}{\mega\hertz}. 
The amplified signal is then transmitted to the modulator through a \SI{10}{\meter} shielded coaxial cable.
The RF chain is directly connected to the bulk lithium-niobate modulator, whose electrical response is predominantly capacitive.
Consequently, the crystal does not provide a matched \SI{50}{\ohm} termination to the RF chain. 
This impedance mismatch produces electrical reflections and, at the same time, contributes to the large voltage excursion reached at the crystal input. 
The configuration therefore allows the high-voltage electrical waveform required for phase modulation to be generated from the low-amplitude AWG signal without a dedicated high-voltage pulse driver. 
The \SI{10}{\meter} cable introduces a controlled propagation delay between the amplifier and the crystal, thereby determining the timing of the reflected signal within the driving chain.
A resistive voltage divider is connected at the crystal input to monitor the applied electrical waveform (LN monitor). 
The divider consists of $R_1=\SI{1}{\kilo\ohm}$ and $R_2=\SI{50}{\ohm}$, providing a monitor voltage $V_{\mathrm{mon}} = 0.0476\,V_{\mathrm{LN}}$, where $V_{\mathrm{LN}}$ is the voltage applied to the crystal.

Since the electrical amplifier is AC coupled, it does not transmit the DC component of the driving waveform. 
To avoid baseline drift, the modulation pulses are therefore generated with zero time average, using bipolar electrical excursions. 
The electrical waveform is synchronized to the optical pulse train such that each optical pulse encounters a well-defined voltage level at the modulator. 
Consequently, successive temporal modes can be assigned different phase shifts by programming the electrical waveform on a pulse-by-pulse basis.

Although the RF amplifier is AC coupled and does not efficiently transmit low-frequency signals, arbitrary slow variations of the phase pattern can still be implemented. 
This is because the modulator is driven by a high-frequency train of bipolar pulses synchronized to the optical pulse sequence, while the desired low-frequency behaviour is encoded in the pulse-to-pulse amplitude variation. 
In spectral terms, a slow modulation produces sidebands around the high-frequency components of the pulse train rather than a large low-frequency electrical component. 
Therefore, programmable phase patterns evolving on timescales much longer than the pulse spacing can be generated while maintaining operation within the amplifier bandwidth.

\subsection{Squeezed-light source and homodyne detection}\label{subsec:squeezed_source}
\begin{figure*}
    \centering
    \includegraphics[width=160mm]{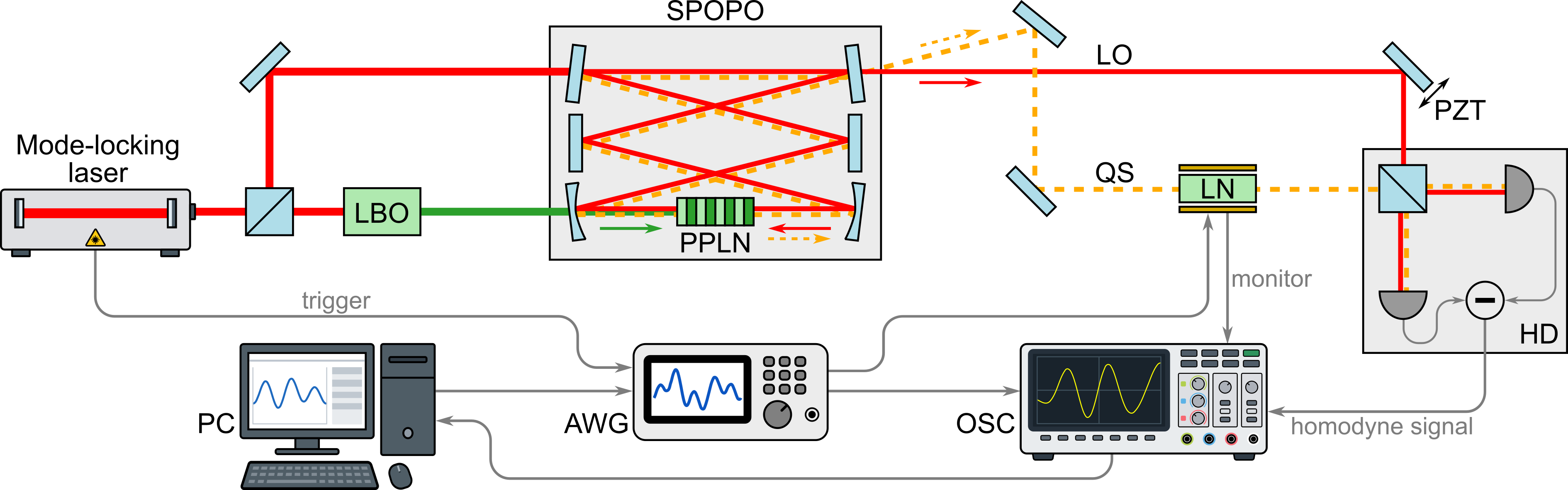}
    \caption{
        Experimental platform for pulse-resolved phase control of squeezed-vacuum states. 
        A mode-locked laser operating at $\SI{1035}{\nano\meter}$ and $\SI{93}{\mega\hertz}$ provides the common optical and timing reference. 
        A fraction of the laser output is frequency doubled in a lithium triborate (LBO) crystal and pumps the periodically poled lithium-niobate (PPLN) crystal inside the synchronously pumped optical parametric oscillator (SPOPO). 
        The SPOPO generates a train of squeezed-vacuum pulses, denoted by QS. 
        The state is transmitted through the bulk lithium-niobate phase modulator (LN) after its generation and is subsequently measured with a time-resolved balanced homodyne detector (HD). 
        The local oscillator (LO) is derived from the counterpropagating cavity-locking field.
        A piezoelectric actuator (PZT) placed along the LO path provides an independent scan of the homodyne phase. 
        The modulation with the AWG and the acquisition with the oscilloscope (OSC) are synchronized to the mode-locked laser. 
    }
    \label{fig:setup}
\end{figure*}

A schematic representation of the complete experimental arrangement is shown in Figure~\ref{fig:setup}.
The quantum states used in the experiment are generated by a synchronously pumped optical parametric oscillator (SPOPO), driven by a mode-locked laser operating at a repetition rate of $\SI{93}{\mega\hertz}$ and a central wavelength of $\SI{1035}{\nano\meter}$.
The optical cavity of the SPOPO primarily provides a well-defined and stable spatial mode for the generated field, thereby simplifying the alignment and mode matching of the source.
The infrared output of the laser is split into two paths. 
One path is frequency-doubled to $\SI{517.5}{\nano\meter}$ with a \SI{5}{\milli\meter}-long lithium triborate (LBO) and used as the pump field for the nonlinear crystal inside the SPOPO, while the other path is injected into the SPOPO and used for cavity stabilization.
The source and its stabilization scheme have been described in detail in previous work~\cite{Suerra2026,Suerra2026_2}.
The SPOPO consists of a six-mirror optical cavity.
At $\SI{1035}{\nano\meter}$, the input and output couplers have reflectivities of $R_\mathrm{ic}=\SI{95}{\percent}$ and $R_\mathrm{oc}=\SI{20}{\percent}$, respectively, while the remaining cavity mirrors are highly reflective.
These mirrors are highly transmissive for the $\SI{517.5}{\nano\meter}$ pump field.
A thermally stabilized $\SI{3}{\milli\meter}$-long periodically poled lithium niobate (PPLN) crystal is used for squeezed-vacuum generation.
The infrared field used for cavity stabilization is coupled into the SPOPO so that it counterpropagates with respect to the quantum field generated by the pump.
The cavity length is actively stabilized using the Pound--Drever--Hall (PDH) technique~\cite{Drever1983}.
After transmission through the cavity, the counterpropagating auxiliary field is used as the local oscillator (LO) for balanced homodyne detection.
This configuration provides a common optical origin for the LO and the quantum field while allowing the stabilization field to propagate in the direction opposite to the generated quantum state.
The optical dispersion introduced by the nonlinear crystal is compensated by one of the cavity mirrors, yielding a net group-delay dispersion close to zero at the operating wavelength.
The resulting squeezed-vacuum field emerging from the SPOPO is therefore sent to the bulk lithium-niobate phase modulator (LN), which is placed downstream of the state-generation process.
The phase transformation is thus applied to the already generated quantum state, independently of the generation process.

The modulated field is characterized using a time-resolved balanced homodyne detector (HD), which resolves the individual pulses of the $\SI{93}{\mega\hertz}$ pulse train, and it is fully described in~\cite{Altilia2026}.
The variance $\sigma^2$ of the quantum state, relative to vacuum noise, measured with the HD, is related to the actual variance at SPOPO output $\sigma^2_\mathrm{SPOPO}$ through the detection efficiency $\eta$ as
\begin{align*}
    \sigma^2 = 1 - \eta\left(1-\sigma_\mathrm{SPOPO}^2\right) \, .
\end{align*}
In this work, the detection efficiency was $\eta=\SI{40}{\percent}$.
This value includes the quantum efficiency of the detector photodiodes, the optical transmission losses and the non-ideal beam splitting, the non-ideal overlap with the local oscillator, and the residual contribution of electronic noise and imperfect common-mode rejection in the balanced detector. 
During the phase-modulation measurements, the LO phase is kept fixed while the phase of the squeezed-vacuum pulses is varied by the electro-optic modulator.
As a reference measurement, the same phase-dependent quadrature variance is independently obtained by varying the LO phase with a piezoelectric actuator (PZT) while leaving the quantum state unchanged.

\section{Device characterization}\label{sec:characterization}
The electro-optic modulator was characterized in two complementary steps.
First, the electrical response at the crystal input was measured to determine the temporal window available for pulse-resolved operation.
Second, the corresponding optical phase shift was calibrated interferometrically, providing the relation between the voltage applied to the crystal and the phase modulation experienced by the optical field.

\subsection{Electrical response}\label{subsec:electrical_response}
The voltage waveform generated by the AWG is amplified and delivered to the lithium-niobate crystal through the $\SI{10}{\meter}$ coaxial cable described in Section~\ref{subsec:modulator}. 
The voltage applied to the crystal is monitored through the resistive divider at the crystal input and recorded with an oscilloscope. 
As the RF amplifier is AC coupled, the driving waveform was designed with zero temporal average. 
Each modulation pulse therefore consists of two consecutive voltage levels of opposite polarity and equal duration.

\begin{figure}
    \centering
    \includegraphics[width=85mm]{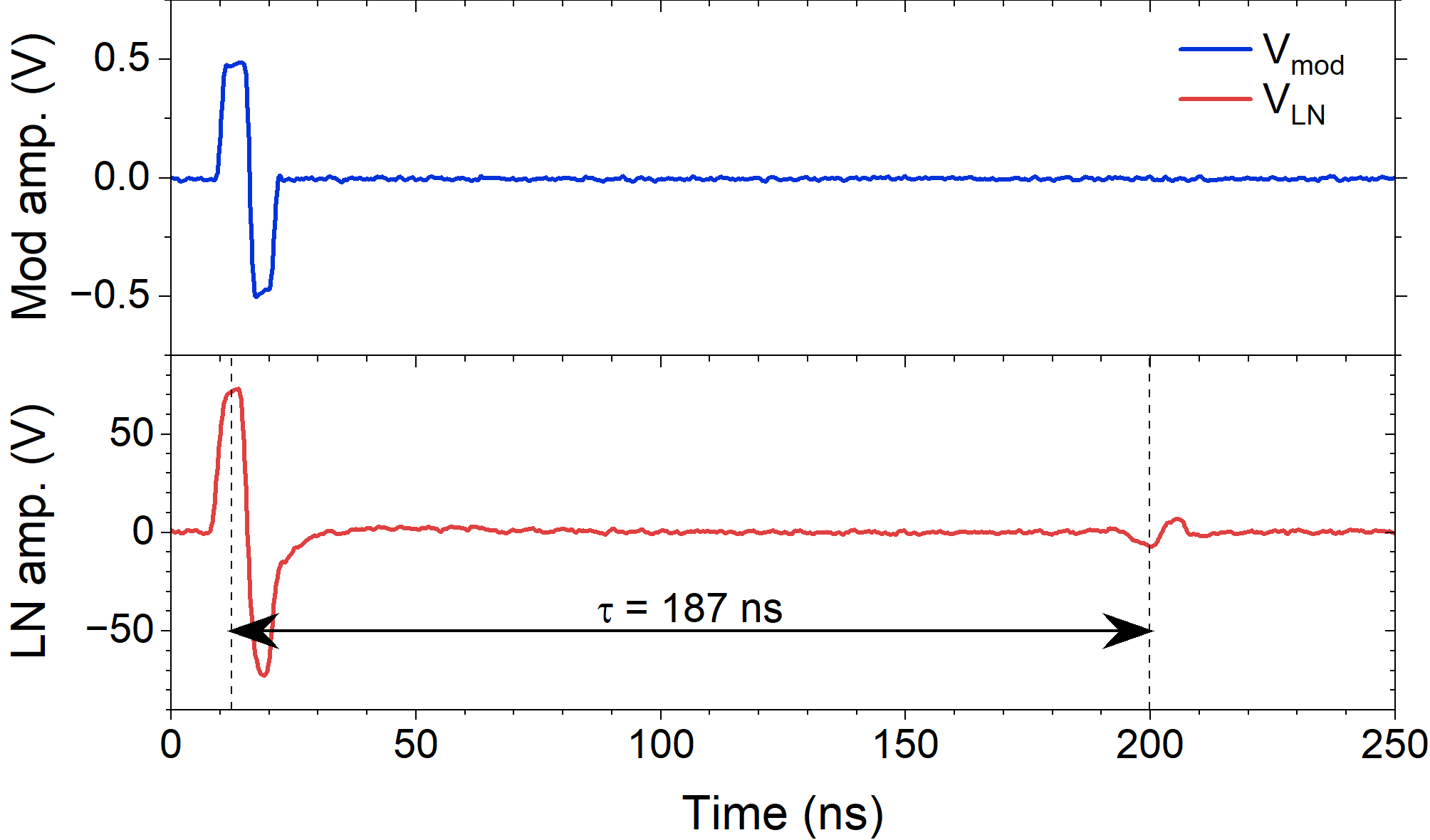}
    \caption{
        Electrical response of the bulk lithium-niobate phase modulator. 
        The upper trace (blue) shows an example of a single modulation pulse generated by the AWG, while the lower trace (red) shows the corresponding voltage measured at the crystal input after RF amplification and propagation through the \SI{10}{\meter} coaxial cable. 
        A delayed reflected transient is observed at $\tau = \SI{187}{\nano\second}$.
    }
    \label{fig:electrical_response}
\end{figure}

Figure~\ref{fig:electrical_response} compares a single pulse programmed at the AWG (top, blue) with the voltage measured at the crystal input through the resistive monitor (bottom, red). 
The measured waveform differs from the programmed signal as a consequence of the finite bandwidth of the electrical chain and of the impedance mismatch between the \SI{50}{\ohm} transmission line and the predominantly capacitive lithium-niobate load.
The crystal capacitance is approximately \SI{10}{\pico\farad}, corresponding to a reactance of about \SI{170}{\ohm} at \SI{93}{\mega\hertz} and \SI{50}{\ohm} at \SI{300}{\mega\hertz}. 
Since the load is therefore strongly reactive over the frequency range relevant to the modulation waveform, it does not provide a matched \SI{50}{\ohm} termination.
As a consequence, a significant fraction of the incident signal is reflected at the crystal. 
The superposition of the incident and reflected waves at the modulator input increases the voltage developed across the crystal, allowing peak-to-peak excursions as large as $\Delta V_\mathrm{LN} \approx \SI{162}{\volt}$.
By comparison, a fully matched \SI{50}{\ohm} termination would substantially reduce the voltage available across the crystal.
A separate effect is responsible for the delayed transient observed at $\tau = \SI{187}{\nano\second}$.
After reflection at the crystal, the electrical pulse propagates back through the \SI{10}{\meter} cable toward the RF amplifier. 
If the amplifier presented an ideal \SI{50}{\ohm} output impedance, the returning wave would be completely absorbed. 
However, the effective output impedance of the amplifier is approximately \SI{39}{\ohm}, resulting in a non-zero reflection coefficient at the source side. 
Consequently, part of the returning pulse is reflected back toward the crystal and reappears as the delayed transient visible in Figure~\ref{fig:electrical_response}. 
The measured delay is consistent with the round-trip propagation time along the cable.
The pulse rise time at the crystal input is $\approx \SI{2.8}{\nano\second}$.
Despite these fast electrical transitions, the voltage excursion at the crystal reaches values of up to approximately \SI{162}{\volt} in the measurements reported below.

In the present implementation, the reflected pulse appearing after \SI{187}{\nano\second} defines a reflection-free temporal window sufficient for $15$ consecutive modulation slots used in the experiment at the \SI{93}{\mega\hertz} repetition rate. 
This sequence length is not an intrinsic limitation of the electro-optic modulator, but is determined by the propagation delay and the uncompensated electrical reflections of the present driving chain. 
Because the AWG can generate arbitrary waveforms, deterministic reflected contributions could in principle be compensated by programmable predistortion, allowing longer sequences to be generated without changing the optical modulator.
A second possibility would be to match the amplifier output impedance to the characteristic impedance of the coaxial line. 
Since the amplifier in the present setup has an output impedance of approximately \SI{39}{\ohm}, adding a \SI{11}{\ohm} series resistor would therefore provide an effective \SI{50}{\ohm} source impedance and eliminate the reflection associated with the source-to-line impedance mismatch.
We experimentally tested this configuration and observed a reduction of the reflected signal, accompanied, however, by a decrease in the voltage delivered to the crystal.
In the final configuration used for the measurements presented in this work, neither of these compensation strategies was implemented, as their optimization and implementation are beyond the scope of the present work.
Finally, at the end of each modulation sequence, a zero-voltage interval is introduced to allow the reflected electrical transients to decay before the subsequent sequence.
This procedure prevents the reflected waveform from overlapping with the modulation slots used for the quantum-state measurements.

\subsection{Interferometric phase calibration}\label{subsec:phase_calibration}
The optical phase response of the modulator was calibrated by interferometry.
A coherent optical field at $\SI{1035}{\nano\meter}$ was transmitted through the lithium-niobate crystal and spatially overlapped with the local oscillator at the homodyne beam splitter.
During this calibration, one photodiode of the balanced detector was blocked, so that the remaining photodiode provided a direct measurement of the interference between the two optical fields.

The relative phase between the coherent field and the local oscillator was first varied using the piezoelectric actuator in the local-oscillator path.
The maximum and minimum detected signals were used to determine the center and amplitude of the interference fringe,
\begin{align*}
    y_{\mathrm{c}} &= \frac{y_{\max}+y_{\min}}{2},\\
    A &= \frac{y_{\max}-y_{\min}}{2} \, .
\end{align*}
Within a monotonic branch of the fringe, the detected signal can then be written as
\begin{align}
    y(\vartheta)=y_\mathrm{c}+A\sin\vartheta \,,
    \label{eq:calibrated_interference_fringe}
\end{align}
where $\vartheta$ denotes the relative optical phase.
The local phase corresponding to a measured signal $y$ is obtained by inverting the fringe relation of Equation~\ref{eq:calibrated_interference_fringe},
\begin{align*}
    \vartheta(y)
    =
    \arcsin\left(
    \frac{y-y_\mathrm{c}}{A}
    \right) \, ,
\end{align*}
with the appropriate branch selected according to the interferometric operating point and the polarity of the applied electrical excursion.
The operating point was chosen near the center of a monotonic fringe, where the interference signal provides high phase sensitivity.
Two electrical excursions with opposite polarity were then applied to the modulator.
For a positive and a negative voltage excursion, corresponding to measured crystal voltages $V_{+}$ and $V_{-}$, respectively, the associated phase shifts were obtained from
\begin{align*}
    \varphi_{+} = \vartheta(y_{+})-\vartheta(y_{0})\,, 
    \quad
    \varphi_{-} = \vartheta(y_{0})-\vartheta(y_{-}) \, ,
\end{align*}
where $y_{0}$ is the signal at the operating point.
The total differential phase excursion is therefore
\begin{equation*}
    \Delta\varphi
    =
    \vartheta(y_{+})-\vartheta(y_{-}) \, .
\end{equation*}

\begin{figure*}
    \centering
    \includegraphics[width=85mm]{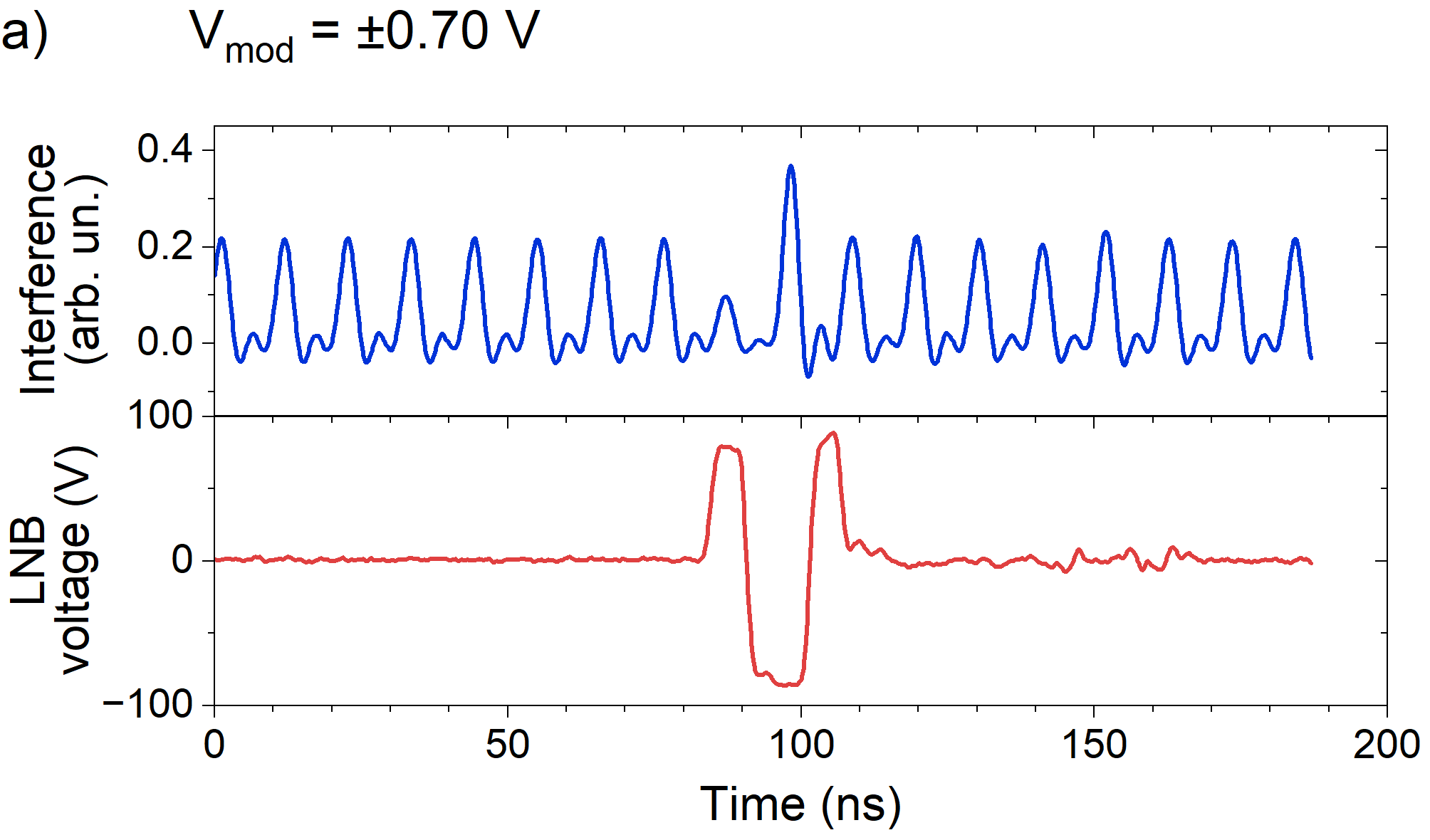}\hspace{5mm}
    \includegraphics[width=85mm]{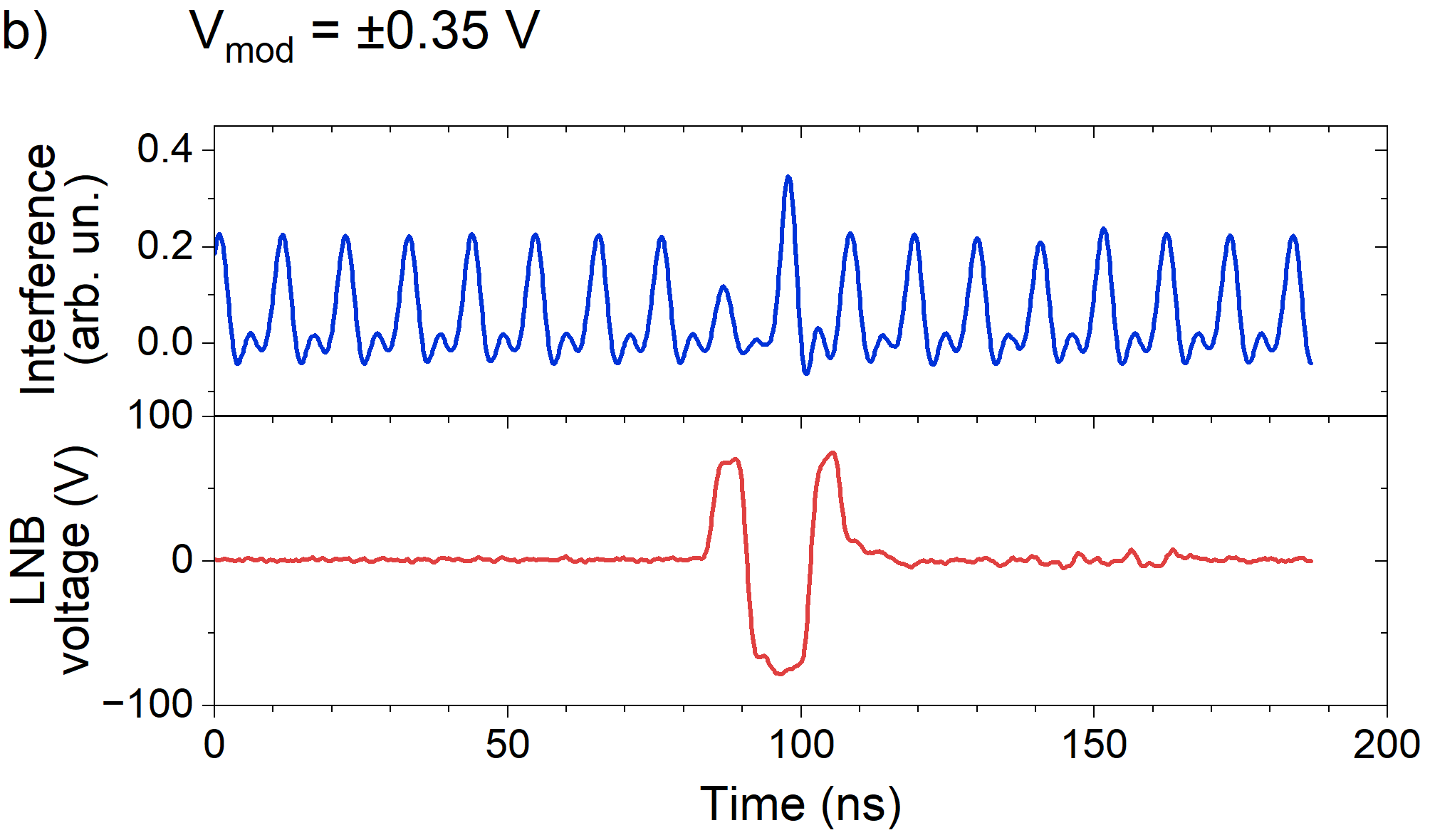}\\
    \vspace{7mm}
    \includegraphics[width=85mm]{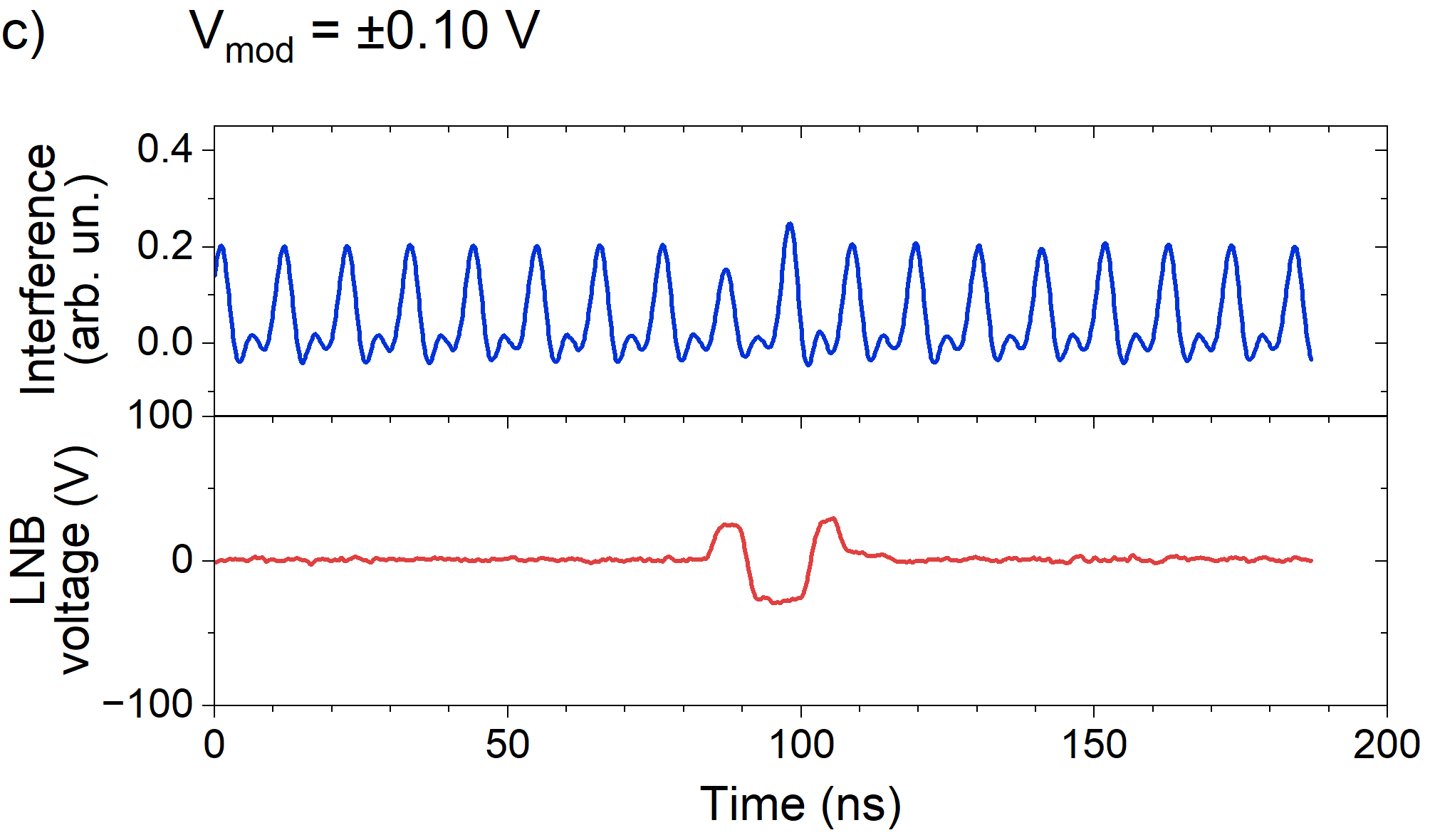}\hspace{5mm}
    \includegraphics[width=85mm]{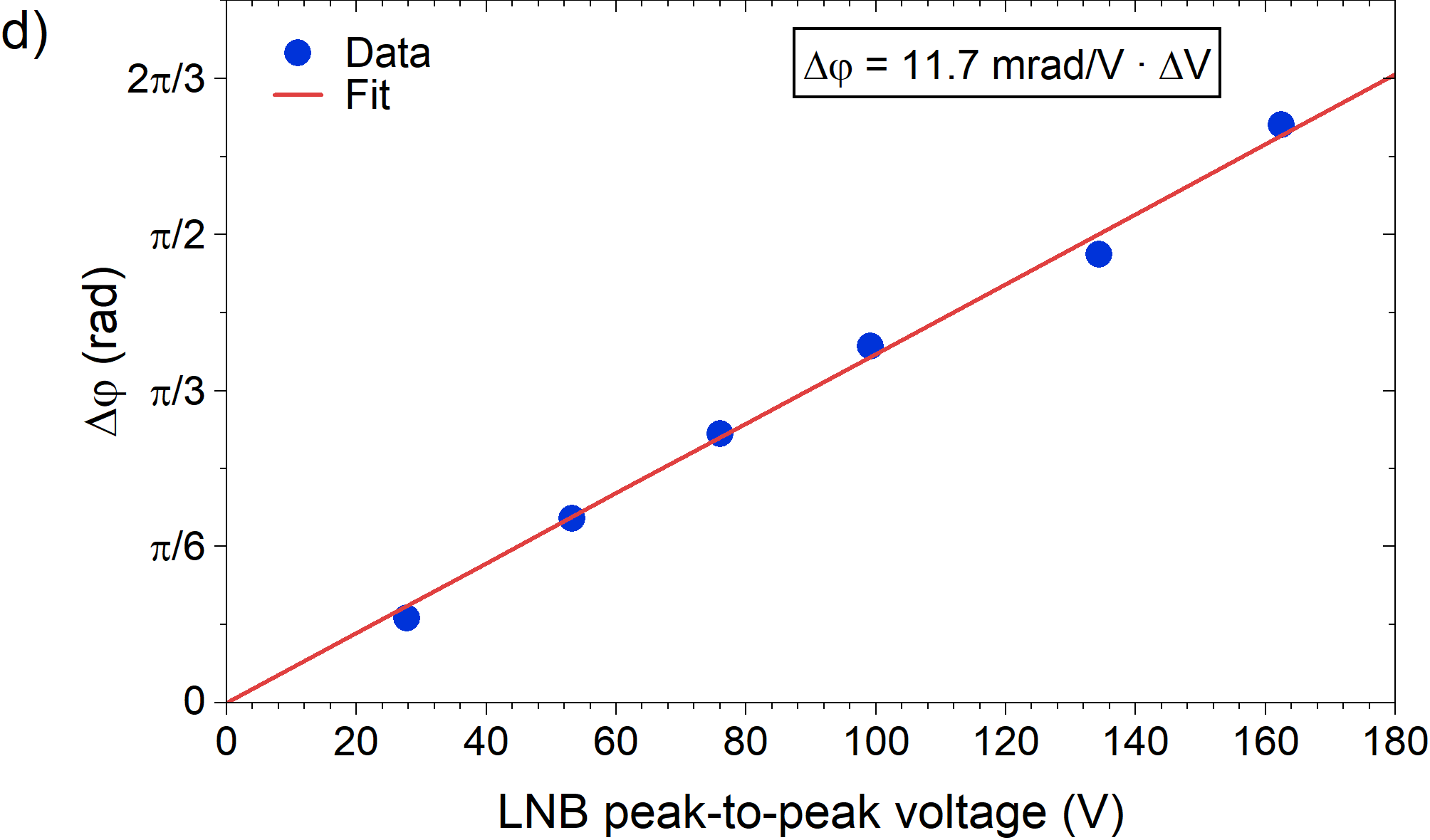}
    \caption{
    Interferometric calibration of the phase shift introduced by the lithium-niobate modulator.
    Panels (a)--(c) show representative electrical modulation pulses (red) and the corresponding optical interference signals (blue) for three different modulation amplitudes.
    The programmed AWG voltages are $\pm\SI{0.70}{\volt}$, $\pm\SI{0.35}{\volt}$, and $\pm\SI{0.10}{\volt}$, respectively.
    Panel (d) reports the measured differential phase shift as a function of the voltage actually applied to the crystal.
    The solid line is a linear fit to the experimental data.
    }
    \label{fig:phasecalibration}
\end{figure*}

Figures~\ref{fig:phasecalibration}a--c show representative electrical waveforms and the corresponding optical interference signals for three different modulation amplitudes.
The experimentally reconstructed phase excursions are shown as a function of the corresponding peak-to-peak voltage $\Delta V_{\mathrm{LN}}$ applied to the crystal in Figure~\ref{fig:phasecalibration}d.
Over the investigated range, the measured response is well described by a linear dependence,
\begin{equation*}
    \Delta\varphi = \alpha \Delta V_{\mathrm{LN}} \,.
\end{equation*}
This calibration directly links the voltage excursion generated by the RF driving chain to the optical phase shift experienced by the modulator, and is used to determine the electrical amplitudes required for the pulse-resolved quantum-state measurements.
A linear fit yields
\begin{equation*}
    \alpha = \SI{11.7}{\milli\radian\per\volt} \, ,
\end{equation*}
corresponding to an effective half-wave voltage of $V_\pi=\pi/\alpha\simeq\SI{269}{\volt}$.
The largest measured differential phase excursion is $\Delta\varphi_{\max}=\SI{1.93}{\radian}$, obtained by feeding the crystal with a differential peak-to-peak voltage on the LN crystal of $\simeq\SI{162}{\volt}$.
This value exceeds $\pi/2$, allowing the modulation to cover both the squeezed and anti-squeezed quadratures of the generated state.
The phase-modulation efficiency could be further improved using a smaller crystal, and thus smaller electrode separation.
For example, $d=\SI{1.5}{\milli\meter}$, a practical value for a bulk implementation, would approximately halve $V_\pi$ for otherwise unchanged geometry. 
Moreover, commercial RF power amplifiers providing up to \SI{100}{\watt} are readily available, which would ideally increase the voltage excursion by a factor of $\sqrt{2}$ with respect to our setup.

The group velocity dispersion of the LN crystal along its extraordinary axis at \SI{1035}{\nano\meter} is $\simeq\SI{243}{\femto\second\squared\per\milli\meter}$, leading to a net dispersion of $\mathrm{GDD}\simeq\SI{9700}{\femto\second\squared}$.
The pulse duration is $\tau \simeq\SI{0.6}{\pico\second}$ FWHM.
The relative temporal broadening due to dispersion can be written as~\cite{Agrawal2013}
\begin{align*}
    \epsilon_{\tau} = 
    \frac{\tau' - \tau}{\tau} = 
    \sqrt{1+\left(\frac{4\ln{2}\,\mathrm{GDD}}{\tau^2}\right)^2} - 1
    \simeq \SI{0.3}{\percent} \, ,
\end{align*}
being $\tau'$ the pulse duration after dispersion.
Since $\epsilon_{\tau}\ll1$, we can consider the dispersion negligible in our conditions.
Furthermore, chirped mirrors could be used for dispersion compensation.

\section{Pulse-resolved phase modulation and data analysis}
\label{sec:methods}
The pulse-resolved phase measurements are performed by synchronously driving the electro-optic modulator and acquiring the corresponding homodyne signal.
The modulation sequence is generated by the AWG and synchronized to the $\SI{93}{\mega\hertz}$ optical pulse train using the laser trigger.
The same timing reference is used throughout the acquisition, ensuring that each electrical modulation pulse is associated with a well-defined optical pulse.

\begin{figure*}
    \centering
    \includegraphics[width=160mm]{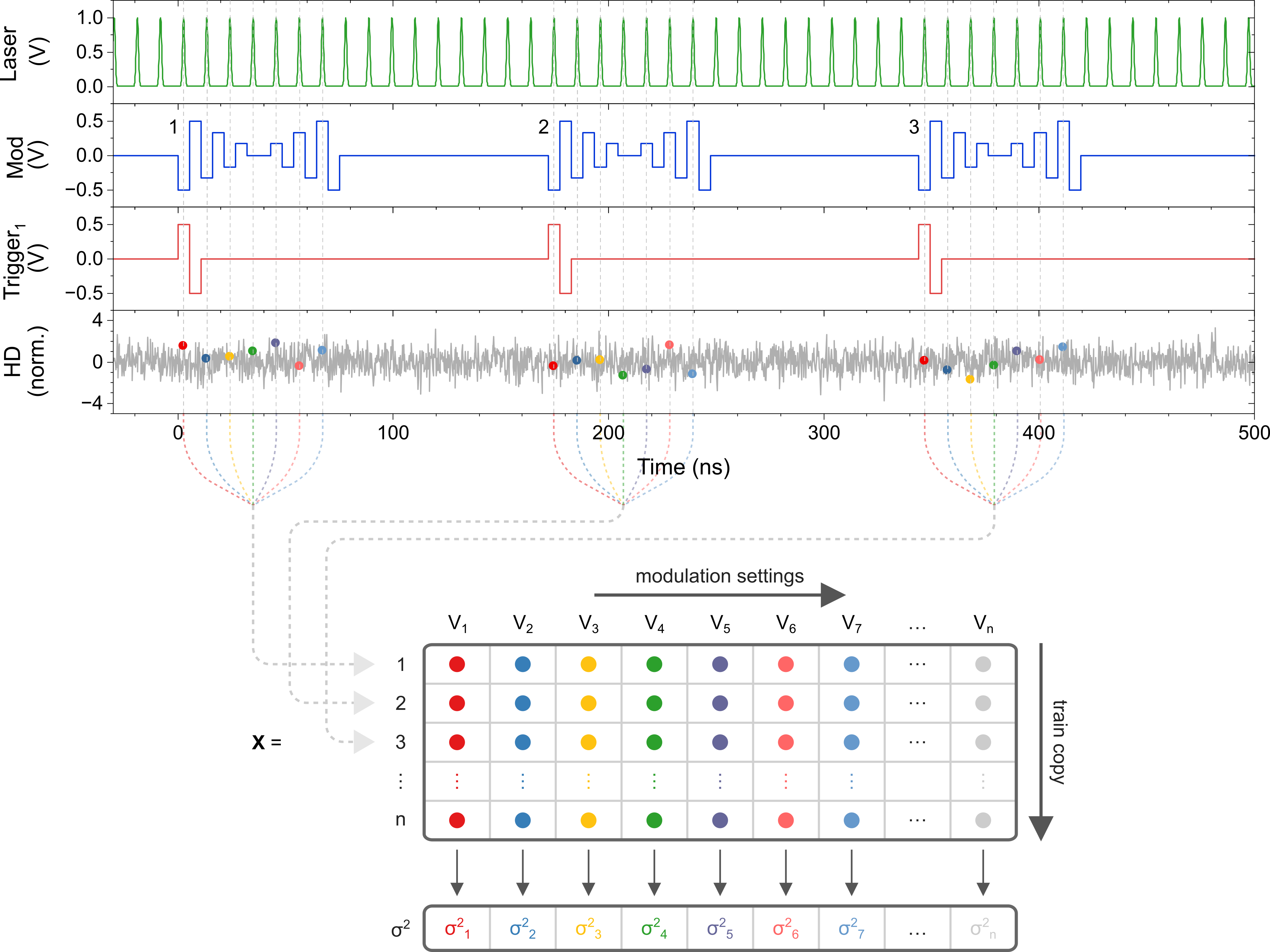}
    \caption{
    Pulse-resolved modulation and data-acquisition scheme.
    The optical pulse train, the electrical modulation waveform, and the synchronization signals are shown together with the corresponding organization of the acquired homodyne data.
    Each modulation pulse is associated with one temporal optical mode.
    The schematic example shows $7$ phase settings for clarity, while the experimental sequence contains $15$ consecutive temporal modes.
    The same sequence of phase settings is repeated $N=2000$ times, allowing the homodyne samples corresponding to each phase setting to be collected over repeated measurements.
    }
    \label{fig:metodimod-matrice}
\end{figure*}

A single modulation sequence, hereafter referred to as a train, consists of $15$ consecutive modulation slots, each corresponding to one optical temporal mode. 
Different modulation amplitudes are programmed for the individual slots, thereby assigning different phase shifts to successive optical pulses.
For a programmed modulation amplitude $V_\mathrm{AWG}$, the bipolar waveform required by the AC-coupled amplifier is generated by applying a voltage $+V_\mathrm{AWG}$ for the first half of the modulation interval and $-V_\mathrm{AWG}$ for the second half.
The resulting waveform has zero time average while providing a well-defined voltage value at the arrival time of each optical pulse.
A second AWG channel provides a synchronization signal associated with the modulation sequence.
For each train, a trigger pulse of $\SI{5}{\nano\second}$ duration is generated at the beginning of the sequence and is used as a temporal reference for the corresponding modulation train.
A third channel generates a single trigger pulse marking the beginning of the complete acquisition sequence.
This signal is used to trigger the oscilloscope, which subsequently records the optical pulse train, the homodyne signal, and the synchronization signal.
Figure~\ref{fig:metodimod-matrice} shows an example of the different signals involved, together with the acquisition rationale described below.
Note that in this illustrative picture a sequence of $7$ phase settings instead of $15$ is reported for simplicity.
The complete train is repeated $N=2000$ times.
The repetition of the same programmed phase sequence allows each temporal position within the train to be sampled repeatedly under identical modulation conditions.
The number of repetitions is chosen as a compromise between sufficient statistical averaging, and the memory and acquisition-rate limitations of the oscilloscope and computer-based acquisition system.

The oscilloscope records the optical pulse train from the laser trigger (green) together with the homodyne signal (gray) and the synchronization reference (blue) over the entire acquisition window.
The temporal positions of the optical pulses are first identified from the synchronization reference and the laser reference.
For each repetition of the modulation sequence, the homodyne signal is then sampled at the same temporal position relative to the corresponding optical pulse (green), as explained in~\cite{Altilia2026}.
In this way, each optical pulse contributes one homodyne sample for a given phase setting, rather than a temporal integral over the pulse.

The selected homodyne data can therefore be arranged into a matrix in which each row corresponds to one repetition of the modulation sequence and each column corresponds to one phase setting, as sketched in the lower panel of Figure~\ref{fig:metodimod-matrice}.
For a train containing $M$ modulation pulses (in our case $15$) and $N$ repetitions (in our case $2000$), the resulting data matrix is
\begin{equation*}
    \mathbf{X}
    =
    \begin{pmatrix}
        x_{1,1} & x_{1,2} & \cdots & x_{1,M}\\
        x_{2,1} & x_{2,2} & \cdots & x_{2,M}\\
        \vdots & \vdots & \ddots & \vdots\\
        x_{N,1} & x_{N,2} & \cdots & x_{N,M}
    \end{pmatrix}\,,
\end{equation*}
where $x_{i,j}$ is the homodyne sample obtained from the $j$-th temporal mode of the $i$-th repetition.
For each phase setting $j$, the corresponding quadrature variance is obtained from the ensemble of $N$ homodyne samples as
\begin{equation}
    \sigma_j^2
    =
    \frac{1}{N-1}
    \sum_{i=1}^{N}
    \left(
        x_{i,j}-\overline{x}_j
    \right)^2,
    \label{eq:quadrature_variance}
\end{equation}
where
\begin{equation*}
    \overline{x}_j
    =
    \frac{1}{N}
    \sum_{i=1}^{N}x_{i,j}
\end{equation*}
is the mean homodyne value for the $j$-th phase setting.
Once normalized to the vacuum variance, the resulting set of variances therefore samples the quadrature noise of the squeezed state at different optical phases while keeping the local-oscillator phase fixed.
Since the phase sequence is repeated identically for every acquisition cycle, the variation of the measured quadrature variance across the columns of $\mathbf{X}$ is attributed to the phase transformation imposed by the electro-optic modulator.

For comparison, an equivalent phase scan is performed by keeping the quantum state unmodulated and varying the local-oscillator phase with the piezoelectric actuator.
In this case, the phase continuously varies over several periods, and the variance is calculated for signal bins of $N = 2000$ elements.

\section{Results}\label{sec:results}
We now demonstrate the operation of the electro-optic modulator on the generated squeezed-vacuum field.
The phase-modulated quantum state is measured with the time-resolved balanced homodyne detector described in Section~\ref{subsec:squeezed_source}.
The phase of the local oscillator is kept fixed throughout the measurement, while the voltage applied to the lithium-niobate crystal is varied from pulse to pulse according to the programmed sequence described in Section~\ref{sec:methods}.
For each modulation setting, the homodyne samples obtained from repeated realizations of the same temporal mode are used to calculate the corresponding quadrature variance according to Equation~\ref{eq:quadrature_variance}.
The measured variance therefore provides a direct measurement of the quadrature noise at the optical phase imposed by the modulator.
The variances are normalized to the shot-noise level measured with the squeezed-vacuum field blocked.

\begin{figure}
    \centering
    \includegraphics[width=85mm]{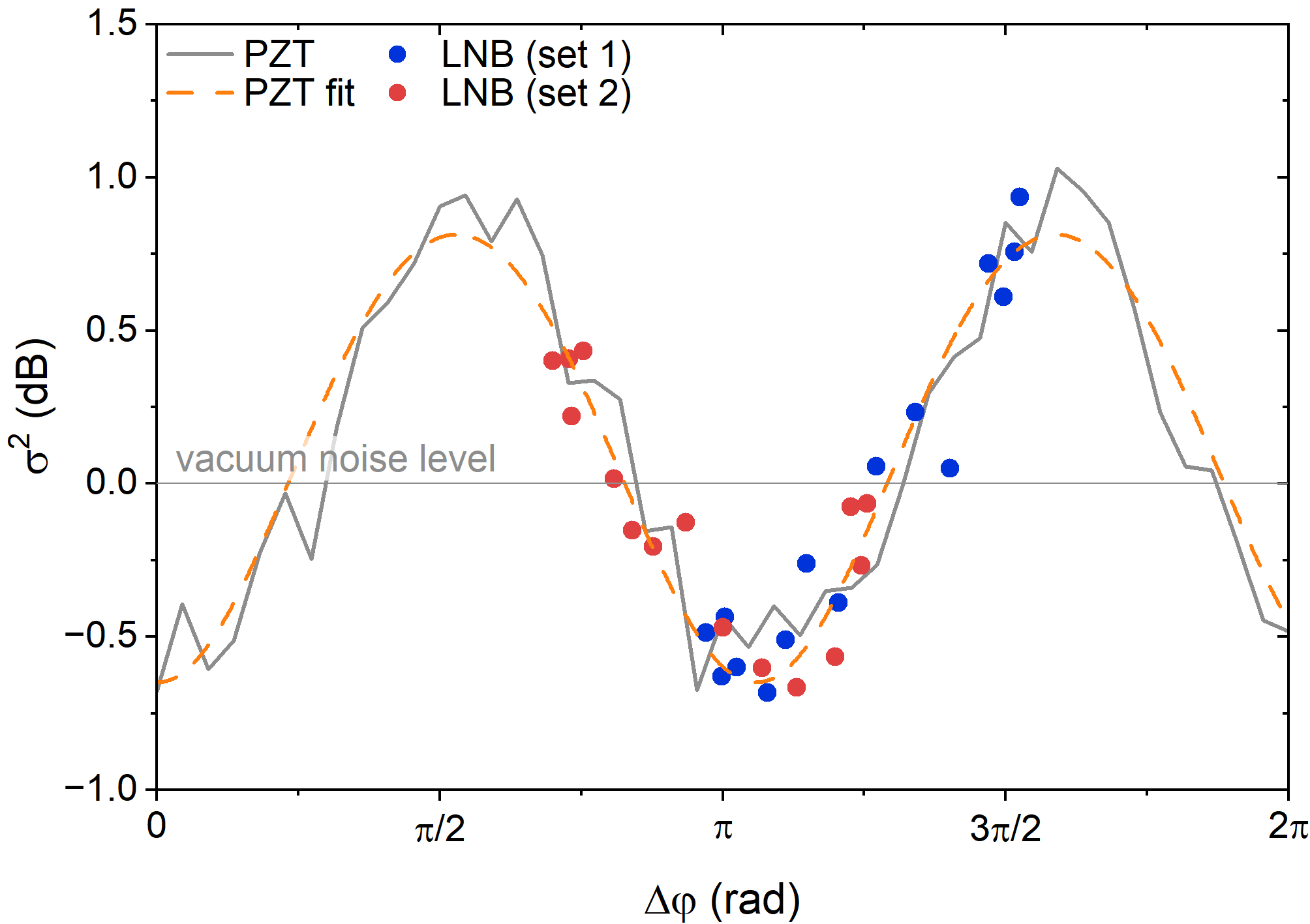}
    \caption{
        Pulse-resolved phase control of the squeezed-vacuum state.
        The quadrature variance $\sigma^2$, normalized to the shot-noise variance, is shown as a function of the relative optical phase $\Delta\varphi$.
        The blue and red markers are two independent pulse-resolved measurements in which the phase of the squeezed-vacuum state is varied by the bulk lithium-niobate electro-optic modulator.
        Each data set contains $15$ consecutive temporal modes, with a different programmed phase assigned to each mode.
        The gray curve is an independent measurement obtained by keeping the quantum state unchanged and scanning the local-oscillator phase with a piezoelectric actuator.
        The orange dashed line is a fit to this latter reference trace.
    }
    \label{fig:pztvslnb}
\end{figure}

Figure~\ref{fig:pztvslnb} compares the quadrature variance obtained by phase modulation of the quantum state (points) with the corresponding reference measurement obtained by scanning the local-oscillator phase (gray line).
In the latter measurement, the quantum state is kept unchanged and the phase of the LO is varied using the piezoelectric actuator.
A fit to the trend obtained from the PZT scan is shown as a dashed orange line.
The fitting function used was~\cite{Olivares2012}
\begin{equation*}
    \sigma^2(\Delta\varphi) =
    a\,\cos^2\!\left(\Delta\varphi-\varphi_0\right) +
    b\,\sin^2\!\left(\Delta\varphi-\varphi_0\right) \, .
\end{equation*}
where $a$ and $b$ are the squeezed- and anti-squeezed-quadrature variances, respectively, and $\varphi_0$ is a reference phase.
The two measurements therefore implement the same effective relative phase scan through two physically different operations: phase rotation of the quantum state and phase rotation of the local oscillator.
Note that we show two different datasets of the pulse-resolved phase control measurement (blue and red), obtained by applying the modulation around different relative phases between LO and quantum state.
The phase-dependent modulation of the quadrature variance follows the expected periodic behavior of a squeezed state.
The measured quadrature variance gives $\approx\SI{-0.65}{\decibel}$ of squeezing, and $\approx\SI{0.9}{\decibel}$ of anti-squeezing relative to the shot-noise level.
After correcting for the detection efficiency $\eta = \SI{40}{\percent}$ introduced in Section~\ref{subsec:squeezed_source}, these values correspond to $\approx\SI{-1.9}{\decibel}$ of squeezing, and $\approx\SI{2.0}{\decibel}$ of anti-squeezing at the SPOPO output.
Importantly, since in the electro-optic measurement the local oscillator remains at a fixed phase throughout the acquisition, the variation of the measured quadrature variance is produced by changing the phase of the quantum state itself.
Moreover, because each modulation setting is assigned to a different temporal position within the $15$-pulse sequence, the phase transformation is applied independently to consecutive optical modes at the $\SI{93}{\mega\hertz}$ repetition rate.

\section{Conclusions}\label{sec:conclusions}
We have demonstrated pulse-resolved post-generation phase control of squeezed-vacuum states at $\SI{1035}{\nano\meter}$, and at a repetition rate of $\SI{93}{\mega\hertz}$ using a low-loss bulk lithium-niobate electro-optic modulator.
The device is driven by a programmable arbitrary waveform generator followed by a broadband commercial RF power amplifier, providing the large voltage excursions required for phase modulation without a dedicated high-voltage pulse driver. 
The driving scheme operates above the piezoelectric-resonance region of lithium niobate and exploits the impedance mismatch with the capacitive crystal load to increase the voltage delivered to the modulator.

An interferometric calibration of the device yielded a maximum measured differential phase excursion of $\SI{1.93}{\radian}$ for a differential crystal peak-to-peak voltage of $\approx\SI{162}{\volt}$.
The voltage-to-phase response is linear over the investigated range, allowing different phase shifts to be programmed for consecutive temporal modes of the $\SI{93}{\mega\hertz}$ optical pulse train. 

Using time-resolved balanced homodyne detection, we measured the phase-dependent quadrature variance of a squeezed-vacuum state. 
By keeping the local-oscillator phase fixed and varying the applied electrical waveform, the phase of successive quantum-state pulses was changed independently at the $\SI{93}{\mega\hertz}$ repetition rate.
The resulting pulse-resolved phase scan agrees with an independent measurement obtained by varying the local-oscillator phase with a piezoelectric actuator, demonstrating pulse-resolved post-generation phase control of the generated quantum state.

Importantly, the demonstrated operation is performed after quantum-state generation.
This separates the generation of the quantum resource from its subsequent temporal-mode-dependent manipulation, providing an electro-optic building block that can be inserted downstream of a common quantum source.
The combination of pulse-resolved operation, low optical loss, and compatibility with bulk free-space optical paths makes the device suitable for applications in time-domain-multiplexed quantum optical systems.
In particular, the demonstrated functionality provides the pulse-resolved phase control required by recirculating optical architectures, where successive temporal modes can encounter the same electro-optic element and undergo programmable phase transformations.
Although the present experiment does not implement the modulator inside a recirculating loop, it experimentally demonstrates the underlying quantum-state manipulation at the required pulse-to-pulse timescale.

\section*{Funding}
This work has been funded by the Istituto Nazionale di Fisica Nucleare (INFN) CSN5 within the project T4QC. 

\section*{Acknowledgments}
The authors acknowledge Ennio Viscione of Servizio Progettazione Meccanica e Officina at INFN LASA.

\bibliography{bibliography}

\end{document}